\documentstyle[preprint,aps,prl,psfig]{revtex}
\tighten
\draft
\begin{document}

\draft

\preprint{\today}
\title{Structural and dynamical properties of sodium silicate 
       melts: An investigation by molecular dynamics computer 
       simulation}
\author{J\"urgen Horbach, Walter Kob, and Kurt Binder}
\address{Institute of Physics, Johannes Gutenberg University,
Staudinger Weg 7, D--55099 Mainz, Germany}
\maketitle

\begin{abstract}
We present the results of large scale computer simulations in which we
investigate the static and dynamic properties of sodium disilicate and
sodium trisilicate melts. We study in detail the static properties of
these systems, namely the coordination numbers, the temperature
dependence of the $Q^{(n)}$ species and the static structure factor,
and compare them with experiments. We show that the structure is
described by a partially destroyed tetrahedral SiO$_4$ network and the
homogeneously distributed sodium atoms which are surrounded on average
by 16 silicon and other sodium atoms as nearest neighbors.  We compare
the diffusion of the ions in the sodium silicate systems with that in
pure silica and show that it is much slower in the latter.  The sodium
diffusion is characterized by an activated hopping through the Si--O
matrix which is frozen with respect to the movement of the sodium
atoms. We identify the elementary diffusion steps for the sodium and
the oxygen diffusion and find that in the case of sodium they are
related to the breaking of a Na--Na bond and in the case of oxygen to
that of a Si--O bond. From the self part of the van Hove correlation
function we recognize that at least two successive diffusion steps of a
sodium atom are spatially highly correlated with each other.  With the
same quantity we show that at low temperatures also the oxygen
diffusion is characterized by activated hopping events.  
\end{abstract}

%\pacs{PACS numbers: 61.20.Lc, 61.20.Ja, 02.70.Ns, 64.70.Pf}

%
%
\section{Introduction}
\label{sec1}
Silicate melts and glasses are an important class of materials in very
different fields, e.g.~in geosciences (since silicates are geologically
the most relevant materials) and in technology (windows, containers,
and optical fibers). From a physical point of view it is a very
challenging task to understand the properties of these materials on a
microscopic level, and in the last twenty years many studies on
different systems have shown that molecular dynamics computer
simulations are a very appropriate tool for this purpose (Angell {\it
et al.}, 1981, Balucani and Zoppi, 1994, Kob, 1999). The main advantage
of such simulations is that they give access to the whole microscopic
information in form of the particle trajectories which of course cannot
be determined in real experiments.

In pure silica (SiO$_2$) the structure is that of a disordered
tetrahedral network in which SiO$_4$ tetrahedra are connected via the
oxygens at the edges. In a recent simulation (Horbach and Kob, 1999a)
we have studied in detail the statics and dynamics of pure silica. In
the present paper we investigate the statics and dynamics of melts in
which the network modifier sodium is added to a silica melt. In
particular two sodium silicate systems are discussed, namely sodium
disilicate (Na$_2$Si$_2$O$_5$) and sodium trisilicate 
(Na$_2$Si$_3$O$_7$) to which will be referred to NS2 and NS3, 
respectively.

In recent years several authors found, by using the potential proposed
by Vessal {\it et al.} (1989) that, e.g., NS2 is characterized by a
microsegregation in which the sodium atoms form clusters of a few atoms
between bridged SiO$_4$ units (Vessal {\it et al.}, 1992, Smith {\it et
al.}, 1995, Cormack and Cao, 1997). This result is somewhat surprising
because in experiments one observes a phase separation between Na$_2$O
and SiO$_2$ at lower temperatures and more likely in sodium silicates
with a higher SiO$_2$ concentration (Mazurin {\it et al.}, 1970, Haller
{\it et al.}, 1974).  In order to see whether microsegregation is also
reproduced with a different model from the one of Vessal {\it et
al.}~we have performed our simulations with a different potential
(discussed below).

Up to now the dynamics of systems like alkali silicates was 
investigated by studying only the dynamics of the alkali atoms
which, at lower temperatures, is much faster than that of the silicon 
and oxygen atoms. In this paper we therefore present the results 
for the diffusion dynamics of all components in NS2 and NS3
as well as the microscopic mechanism of diffusion in these systems.

\section{Model and details of the simulations}
\label{sec2}
In a classical molecular dynamics computer simulation one solves
numerically Newton's equations of motion for a many particle system. If
quantum mechanical effects can be neglected such simulations are able
to give in principle a realistic description of any molecular system.
The determining factor of how well the properties of a real material
are reproduced by a MD simulation is the potential with which
the interaction between the atoms is described.  The model potential we
use to compute the interaction between the ions in sodium silicates is
the one proposed by Kramer {\it et al.}~(1991) which is a
generalization of the so called BKS potential (van Beest {\it et al.},
1990) for pure silica. It has the following functional form:
\begin{equation}
\phi(r)=
\frac{q_{\alpha} q_{\beta} e^2}{r} + 
A_{\alpha \beta} \exp\left(-B_{\alpha \beta}r\right) -
\frac{C_{\alpha \beta}}{r^6}\quad \alpha, \beta \in
[{\rm Si}, {\rm Na}, {\rm O}].
\label{eq1}
\end{equation}
Here $r$ is the distance between an ion of type $\alpha$ and an ion of
type $\beta$. The values of the parameters $A_{\alpha \beta}, B_{\alpha
\beta}$ and $C_{\alpha \beta}$ can be found in the original
publication. The potential (\ref{eq1}) has been optimized by Kramer
{\it et al.}~for zeolites, i.e.~for systems that have Al ions in 
addition to Si, Na and O. In that paper the authors used for silicon
and oxygen the {\it partial} charges $q_{{\rm Si}}=2.4$ and 
$q_{{\rm O}}=-1.2$, respectively, whereas sodium was assigned its real
ion charge $q_{{\rm Na}}=1.0$. With this choice charge neutrality is
not fulfilled in sodium silicate systems. To overcome this problem we
introduced for the sodium ions a distance dependent charge $q(r)$
instead of $q_{{\rm Na}}$,
\begin{equation}
  q(r)= \left\{
    \begin{array}{l@{\quad \quad}l}
       0.6 \left( 1+
     \ln \left[ C \left(r_{{\rm c}}-r \right)^2+1 \right] \right) & 
        r < r_{{\rm c}} \\
      0.6  & r \geq r_{{\rm c}}
    \end{array} \right. \label{eq2}
\end{equation}
which means that for $r \ge r_{{\rm c}}$ charge neutrality is valid
($q(r)=0.6$ for $r \ge r_{{\rm c}}$). Note that $q(r)$ is continuous at
$r_{{\rm c}}$. We have fixed the parameters $r_{{\rm c}}$ and $C$ such
that the experimental mass density of NS2 and the static 
structure factor from a neutron scattering experiment are reproduced
well. From this fitting we have obtained the values $r_{{\rm
c}}=4.9$~\AA~and $C=0.0926$~\AA$^{-2}$. With this choice the charge
$q(r)$ crosses smoothly over from $q(r)=1.0$ 
at $1.7$ \AA~to $q(r)=0.6$ for $r \ge r_{{\rm c}}$.

The simulations have been done at constant volume with the density of
the system fixed to $2.37 \, {\rm g}/{\rm cm}^3$. The systems consist
of $8064$ and $8016$ ions for NS2 and NS3, respectively. The reason for
using such a relatively large system size is that, like in SiO$_2$
(Horbach {\it et al.}, 1996, Horbach {\it et al.}, 1999b), strong
finite size effects are present in the dynamics of smaller systems
which have to be avoided (Horbach and Kob, 1999c). The equations of
motion were integrated with the velocity form of the Verlet algorithm
and the Coulombic contributions to the potential and the forces were
calculated via Ewald summation. The time step of the integration was
$1.6$~fs. The temperatures investigated are $4000$~K, $3400$~K,
$3000$~K, $2750$~K, $2500$~K, $2300$~K, $2100$~K, and in addition for
NS2 also $1900$~K.  The temperature of the system was controlled by
coupling it to a stochastic heat bath, i.e.~by substituting
periodically the velocities of the particles with the ones from a
Maxwell-Boltzmann distribution with the correct temperature.  After the
system was equilibrated at the target temperature, we continued the run
in the microcanonical ensemble, i.e.~the heat bath was switched off. In
order to improve the statistics we have done two independent runs at
each temperature.  The production runs have been up to $7.5$~ns real
time which corresponds to $4.5$ million time steps. In the temperature
range under investigation the pressure decreases monotonically from
$3.8$~GPa to $1.8$~GPa in the case of NS3 and from $7.3$~GPa to
$4.1$~GPa in the case of NS2.  In order to make a comparison of the
static structure factor for NS2 from our simulation with one from a
neutron scattering experiment (see below) we have also determined the
structures of the glass at $T=300$~K. The glass state was produced by
cooling the system from equilibrated configurations at $T=1900$~K with
a cooling rate of $1.16 \cdot 10^{12} \, {\rm K}/{\rm s}$. The pressure
of the system at $T=300$~K is $0.96$~GPa.

In the following we compare the properties of NS2 and NS3 with
those of pure silica which we have investigated in recent simulation.
The details of the latter can be found in Horbach and Kob (1999a).
We only mention here that these simulations were done also at the 
density $2.37 {\rm g/cm}^3$ in the temperature range 
$6100 \; {\rm K} \ge T \ge 2750 \; {\rm K}$ for a system of 8016
particles. The two production runs at the lowest temperature were
over about 20 ns real time, and the pressure at this temperature is
$0.9$~GPa.

\section{Results}
\label{sec3}
\subsection{Structural properties}
\begin{figure}[h]
\psfig{file=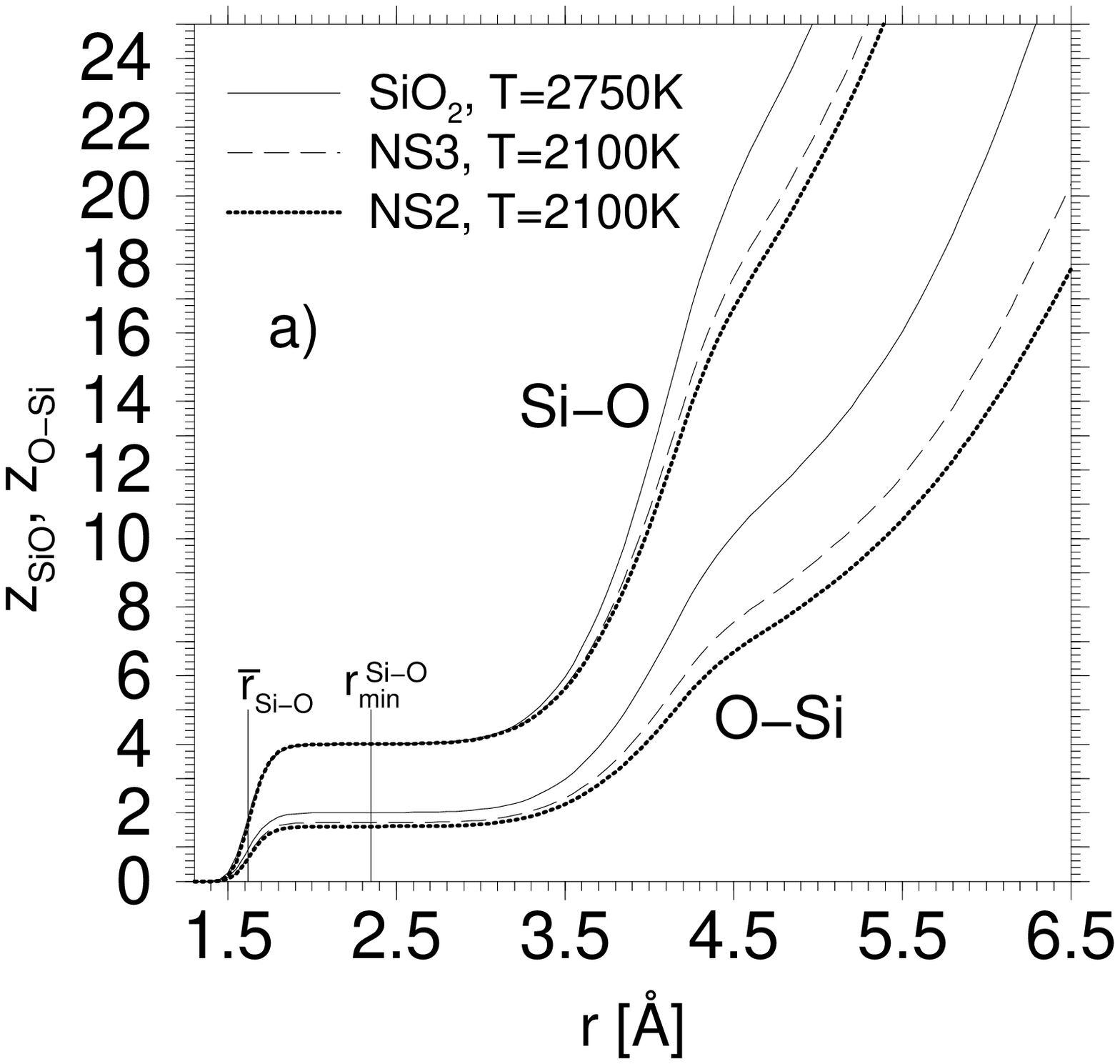,width=13cm,height=9.0cm}
\psfig{file=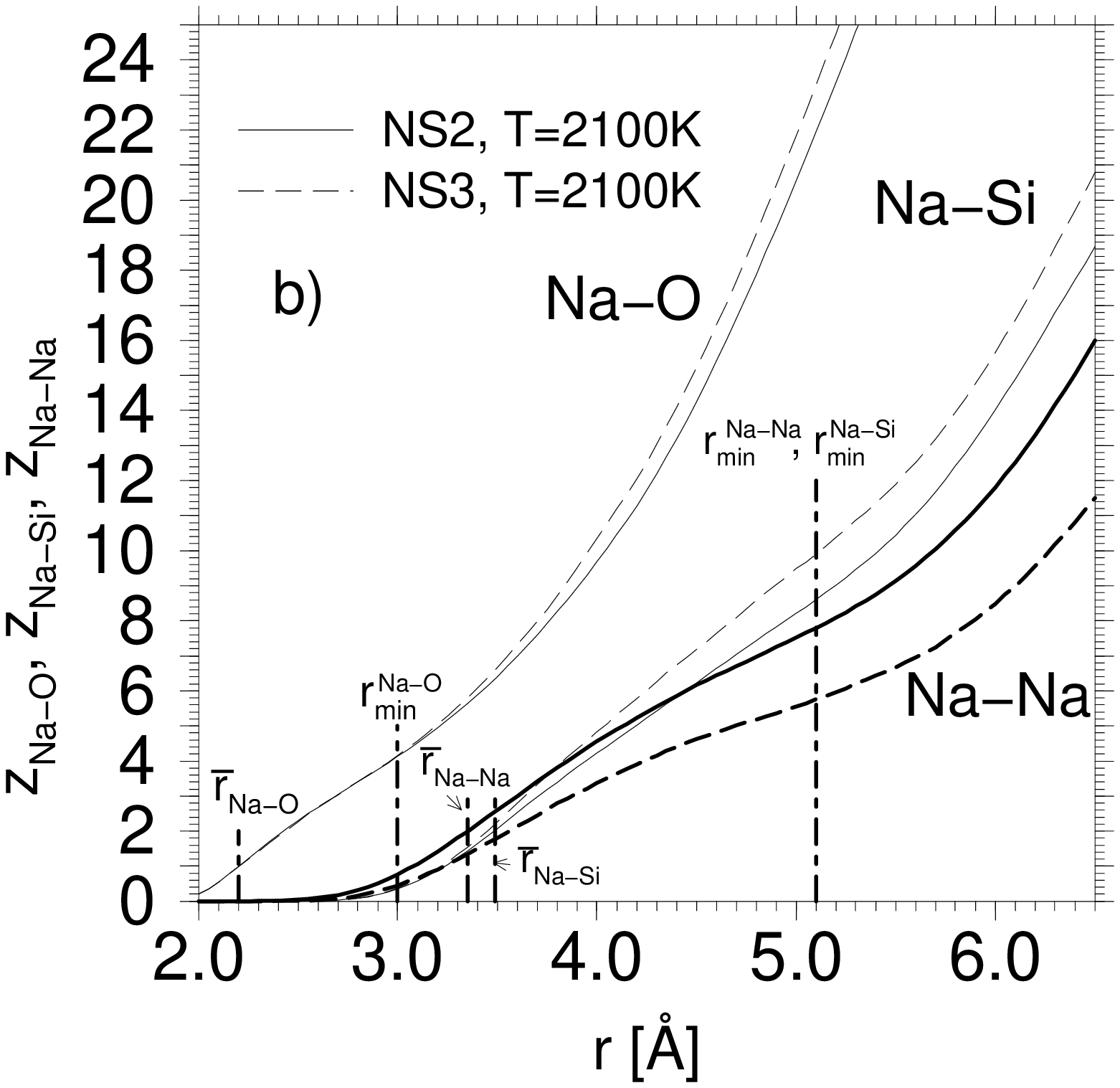,width=13cm,height=9.0cm}
\caption{Coordination numbers $z_{\alpha \beta}$ as a function of 
         distance $r$ at the temperature $T=2750$ K for SiO$_2$ and
         at $T=2100$ K for NS2 and NS3. a) $z_{{\rm Si-O}}(r)$ and
         $z_{{\rm O-Si}}(r)$, b) $z_{{\rm Na-O}}(r)$, 
         $z_{{\rm Na-Si}}(r)$, and $z_{{\rm Na-Na}}(r)$. For the 
         explanation of the vertical lines see text.}
\label{fig1}
\end{figure}

In order to investigate the local environment of an atom in a
disordered structure, especially at high temperatures, it is useful to
calculate the coordination number $z_{\alpha \beta}(r)$ which gives the
number of atoms of type $\beta \in [{\rm Si,Na,O}]$ surrounding an atom
of type $\alpha \in [{\rm Si,Na,O}]$ within a distance $r^{\prime} \le
r$. Note that $z_{\alpha \beta}(r)$ is essentially the integral from
$r^{\prime} = 0$ to $r^{\prime} = r$ over the function $4 \pi r^{\prime
2} g_{\alpha \beta}(r^{\prime})$, where $g_{\alpha \beta}(r)$ denotes
the pair correlation function for $\alpha \beta$ correlations (Balucani
and Zoppi, 1994).  In Fig.~\ref{fig1}a we show $z_{{\rm Si-O}}(r)$ and
$z_{{\rm O-Si}}(r)$ for SiO$_2$ at $T=2750$~K and for NS3 and NS2 at
$T=2100$~K. In both quantities we observe a strong increase 
for distances from $1.5$
\AA~to $1.7$ \AA~at which the functions reach a plateau which persists
up to about $3.0$~\AA.  For distances $r> 3.0$ \AA~there is no such
step like behavior because of the disorder.  The vertical lines in
Fig.~\ref{fig1}a at $\bar{r}_{{\rm Si-O}}=1.61$~\AA~and $r_{{\rm
min}}^{{\rm Si-O}}=2.35$~\AA~correspond to the first maximum and the
first minimum in $g_{{\rm SiO}}(r)$, respectively.  In $z_{{\rm
Si-O}}(r)$ the plateau is essentially at $z=4$ in SiO$_2$ as well as in
the sodium silicate systems which means that most of the silicon atoms,
i.e.~more than 99~\%, are surrounded by four oxygen atoms thus forming
a tetrahedron. In the case of SiO$_2$ there is also a plateau
essentially at $z_{{\rm O-Si}}=2$ in the same $r$ interval. Therefore,
in this system most of the the oxygen atoms are bridging oxygens
between two tetrahedra, and thus the system seems to form a perfect
disordered tetrahedral network even at the relatively high temperature
$T=2750$~K.  Indeed we found that at this temperature less than one
percent of the silicon and oxygen atoms are defects (Horbach and Kob,
1999a).  For NS3 and NS2 a plateau is formed at $z_{{\rm O-Si}}=1.71$
and $z_{{\rm O-Si}}=1.6$, respectively. This is due to the fact that
only a part of the oxygen atoms are bridging oxygens, namely $68.5$ \%
in the case of NS2 and $71.3$ \% in the case of NS3. Most of the other
oxygen atoms form dangling bonds with one silicon neighbor (28.4~\% in
NS2 and 22.7~\% in NS3 at $r_{{\rm min}}^{{\rm Si-O}}$) or they are not
nearest neighbors of silicon atoms (3.1~\% in NS2 and 6.0~\% in NS3 at
$r_{{\rm min}}^{{\rm Si-O}}$).  The local environment of the sodium
atoms is characterized by $z_{{\rm Na-O}}(r)$, $z_{{\rm Na-Si}}(r)$,
and $z_{{\rm Na-Na}}(r)$ in Fig.~\ref{fig1}b.  In these quantities
there is no step like behavior also at small distances. At $r_{{\rm
min}}^{{\rm Na-O}}$, which is approximately at $r=3.0$ \AA~for NS3 and
NS2, the apparent coordination number $z=4.2$ is too small in
comparison with X--ray experiments from which one would expect a value
between 5 and 6 (see Brown {\it et al.}, 1995, and references
therein).  Also $\bar{r}_{{\rm Na-O}}=2.2$ \AA~is too small in
comparison with the values found in X--ray experiments which are
between 2.3~\AA~and 2.6~\AA.  We recognize in Fig.~\ref{fig1}b that in
the case of NS2 the functions $z_{{\rm Na-Na}}(r)$ and $z_{{\rm
Na-Si}}(r)$ are relatively close to each other for $r \le r_{{\rm
min}}^{{\rm Na-Na}}, r_{{\rm min}}^{{\rm Na-Si}}\approx 5.1$~\AA.  So
at $r=r_{{\rm min}}^{{\rm Na-Na}}$ and $r=r_{{\rm min}}^{{\rm Na-Si}}$
the apparent coordination numbers are $z_{{\rm Na-Na}}=7.8$ and
$z_{{\rm Na-Si}}=8.6$, respectively, i.e., are quite similar.  In
contrast to that, the coordination numbers in NS3 at $r=r_{{\rm
min}}^{{\rm Na-Na}}$ and $r=r_{{\rm min}}^{{\rm Na-Si}}$ are $z_{{\rm
Na-Na}}=5.8$ and $z_{{\rm Na-Si}}(r)=9.8$, respectively, which means
that there is a substitution effect in NS3 such that more silicon atoms
are on average in the neighborhood of an sodium atom than in NS2
because there are less sodium atoms in NS3.  Therefore, in NS2 as well
as in NS3 every sodium atom has on average about 16 silicon and sodium
atoms in its neighborhood.

\begin{figure}[h]
\psfig{file=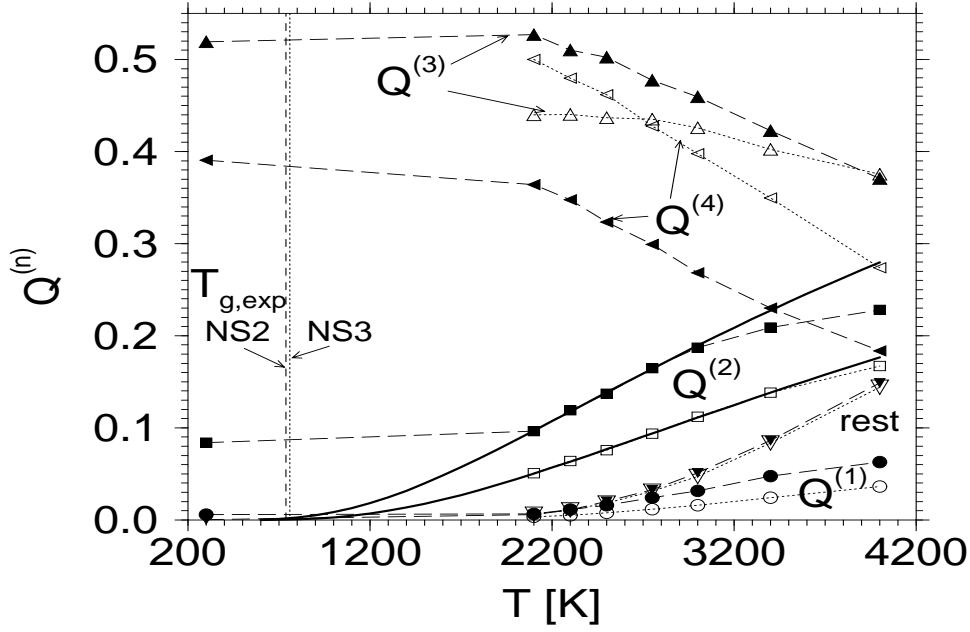,width=13cm,height=9.0cm}
\caption{Temperature dependence of the $Q^{(n)}$ species for 
         NS2 (closed symbols) and NS3 (open symbols). The bold
         lines are fits with Arrhenius laws with the activation 
         energies $E_{{\rm A}}=5441$~K for NS2 and 
         $E_{{\rm A}}=6431$~K for NS3.}
\label{fig2}
\end{figure}

The structure of NS2 and NS3 can be studied in more detail by looking
at the so called $Q^{(n)}$ species which can be determined
experimentally by NMR (Stebbins, 1995) and by Raman spectroscopy
(McMillan and Wolf, 1995).  $Q^{(n)}$ is defined as the fraction of
SiO$_4$ tetrahedra with $n$ bridging oxygens in the system. In
Fig.~\ref{fig1}a we have seen that these structural elements are very
well defined also at temperatures as high as $T=2100$~K. From
Fig.~\ref{fig2} we recognize that at $T=2100$ K there is essentially a
ternary distribution of $Q^{(2)}$, $Q^{(3)}$, and $Q^{(4)}$ both in NS2
and in NS3. At higher temperatures there is also a significant
contribution of silicon defects, i.e.~three-- and five--fold
coordinated silicon atoms (denoted as rest in the figure), and of
$Q^{(1)}$ species. For $T \le 3200$ K the curves for $Q^{(2)}$ are well
described by Arrhenius laws $f(T) \propto \exp(E_{{\rm A}}/T)$ (bold
solid lines in Fig.~\ref{fig2}) with activation energies $E_{{\rm
A}}=5441$ K and $E_{{\rm A}}=6431$ K for NS2 and NS3, respectively. If
we extrapolate these Arrhenius laws to low temperatures we recognize
from Fig.~\ref{fig2} that the $Q^{(2)}$ species essentially disappear
slightly above the experimental glass transition temperatures $T_{{\rm
g, exp}}$ for NS2 and NS3, which are at 740 K and 760 K (Knoche {\it et
al.}, 1994), respectively.  Therefore, if we would be able to
equilibrate our systems down to $T_{{\rm g, exp}}$ we would expect a
binary distribution of $Q^{(3)}$ and $Q^{(4)}$ species in the glass
state of NS2 and NS3.  Also included in Fig.~\ref{fig2} is the
$Q^{(n)}$ species distribution for NS2 at $T=300$ K which we have
calculated from the configurations which we have produced by cooling
down the system from $T=1900$~K to $T=300$~K with a cooling rate of
$10^{12}$~${\rm K/s}$. We recognize from the data that the $Q^{(n)}$
species distribution essentially coincides at $T=300$ K with the one at
$T=2100$~K, which is due to the fact that the liquid structure at the
latter temperature has just been frozen in.  If we extrapolate the data
from $2100$ K to lower temperatures allowing further relaxation we
would expect about 45 \% $Q^{(3)}$ and 55 \% $Q^{(4)}$ in the case of
NS3 and about 40 \% $Q^{(3)}$ and 60 \% $Q^{(4)}$ in the case of NS2 at
$T_{{\rm g, exp}}$.  In contrast to these values one finds in NMR and
Raman experiments 60 \% $Q^{(3)}$ and 40 \% $Q^{(4)}$ in the case of
NS3 and 92 \% $Q^{(3)}$ and 8 \% $Q^{(4)}$ in the case of NS2
(Stebbins, 1988, Mysen and Frantz, 1992, Sprenger {\it et al.}, 1992,
Knoche, 1993).  Thus our simple model is not able to describe the
$Q^{(n)}$ species reliably. The reason for this is probably that our
model gives a slightly wrong coordination function $z_{{\rm Na-O}}(r)$
as we have mentioned before.

\begin{figure}[h]
\psfig{file=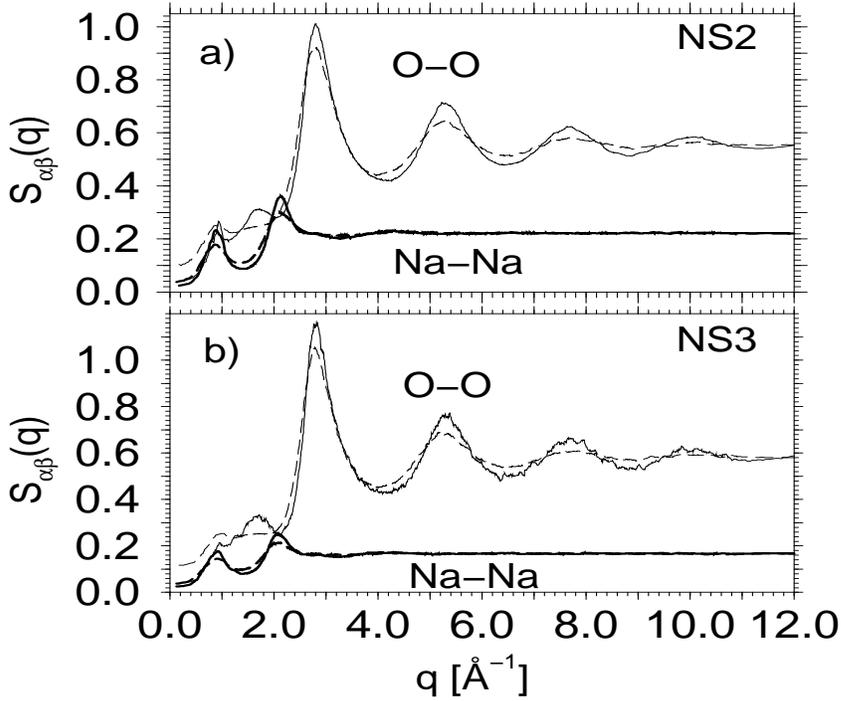,width=13cm,height=9.0cm}
\caption{Partial static structure factors $S_{{\rm OO}}(q)$ (thin
	 lines) and $S_{{\rm NaNa}}(q)$ (bold lines) at the
         temperatures $T=4000$~K (dashed lines) and $T=2100$~K
         (solid lines), a) NS2, b) NS3.} 
\label{fig3}
\end{figure}

So far we have looked at the local environment of the atoms. In order
to investigate the structure on a larger length scale useful 
quantities are the partial static structure factors,
\begin{equation}
  S_{\alpha \beta}(q) = \frac{f_{\alpha \beta}}{N}
                        \sum_{l=1}^{N_{\alpha}}
                        \sum_{m=1}^{N_{\beta}} \;
                \left< \exp( i {\bf q} \cdot ({\bf r}_l - 
		    {\bf r}_m )) \right> , \label{eq3}
\end{equation}
depending on the magnitude of the wave--vector ${\bf q}$.  The factor
$f_{\alpha \beta}$ is equal to 0.5 for $\alpha \ne \beta$ and equal to
1.0 for $\alpha = \beta$. As examples Figs.~\ref{fig3}a and \ref{fig3}b
show $S_{{\rm OO}}(q)$ and $S_{{\rm NaNa}}(q)$ at the temperatures
$T=4000$~K and $T=2100$~K for NS2 and NS3, respectively.  The peak at
$2.8$~\AA$^{-1}$ in $S_{{\rm OO}}(q)$ for NS2 and NS3 corresponds to
the length scale $2 \pi / 2.8$~\AA$^{-1} = 2.24$~\AA~which is
approximately the period of the oscillations in $g_{{\rm OO}}(r)$. In
the same way the peak at $2.1$~\AA$^{-1}$ in $S_{{\rm NaNa}}(q)$ is due
to the period of oscillations in $g_{{\rm NaNa}}(r)$. Moreover, the
temperature dependence we observe for the peaks at $2.8$~\AA$^{-1}$ and
$2.1$~\AA$^{-1}$ is relatively weak.  At $T=2100$~K a peak is visible
at $1.7$~\AA$^{-1}$ in $S_{{\rm OO}}(q)$ for NS2 and NS3, which is the
so called first sharp diffraction peak (FSDP). This feature arises from
the tetrahedral network structure since the length scale which
corresponds to it, i.e.~$2 \pi / 1.7 \; {\rm \AA}^{-1} = 3.7 \; {\rm
\AA}$, is approximately the spatial extent of two connected SiO$_4$
tetrahedra.  Only a shoulder can be identified in $S_{{\rm OO}}(q)$ for
$T=4000$~K at the position of the FSDP because the network structure is
less pronounced than at $T=2100$~K. In agreement with the
aforementioned interpretation of the FSDP this feature is absent in the
Na--Na correlations. We note that this holds also for the Na--O and
Si--Na correlations.  We see therefore that the structure in NS2 is
characterized by a partially destroyed tetrahedral network, and that on
the other hand there are the sodium atoms which are homogeneously
distributed in the system having on average 16 sodium and silicon
neighbors. We would therefore expect that there is a characteristic
length scale of regions where the network is destroyed and where the
sodium atoms are located.  This explains the peak at
$q=0.95$~\AA$^{-1}$ in the $S_{\alpha \beta}(q)$ corresponding to a
length scale $2 \pi/0.95 \; {\rm \AA}^{-1}=6.6$~\AA~which is
approximately two times the mean distance of nearest Na--Na or Na--Si
neighbors. We mention that this peak is also present in the Si--Si,
Si--O, Si--Na, and Na--O correlations. We emphasize that no evidence
for a microsegregation is found in the partial structure factors both
for NS2 and for NS3, at least at the density of this study $2.37 \;
{\rm g/cm}^3$.

\begin{figure}[h]
\psfig{file=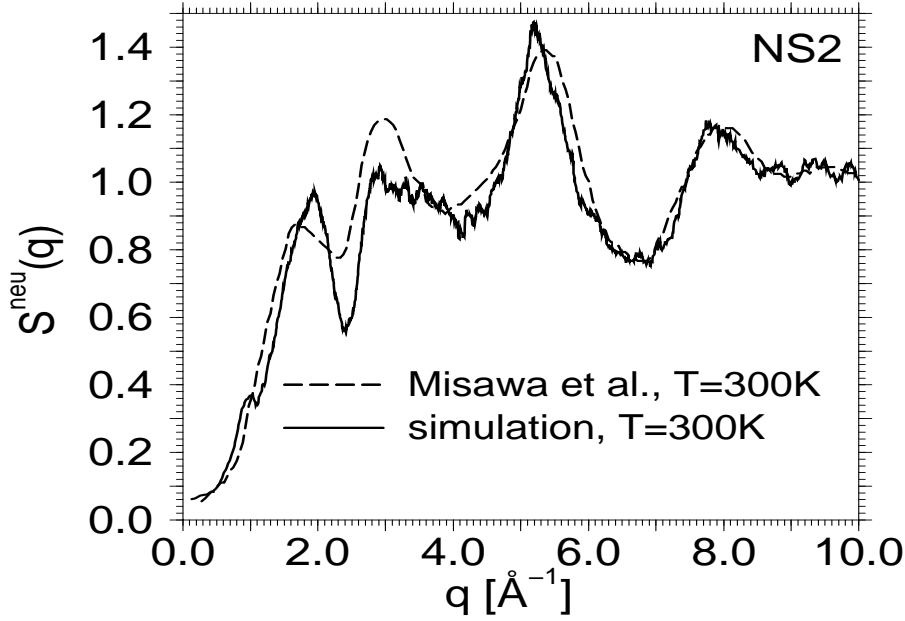,width=13cm,height=9.0cm}
\caption{Comparison of the static structure factor $S^{{\rm neu}}(q)$
         from our simulation (solid line) with the experimental
         data of Misawa {\it et al.}~(1980)
         (dashed line).}
\label{fig4}
\end{figure}

In order to make a comparison of the static structure factor for NS2
measured in a neutron scattering experiment by Misawa {\it et al.}
(1980) with that from our model we have to determine $S^{{\rm neu}}(q)$
which is calculated by weighting the partial structure factors from the
simulation with the experimental coherent neutron scattering lengths
$b_{\alpha}$ ($\alpha \in [{\rm Si, Na, O}]$):
\begin{equation}
   S^{{\rm neu}}(q) = 
     \frac{1}{\sum_{\alpha} N_{\alpha} b_{\alpha}^2}
     \sum_{\alpha \beta} b_{\alpha} b_{\beta} S_{\alpha \beta}(q)
     .  \label{eq4}
\end{equation}
The values for $b_{\alpha}$ are $0.4149 \cdot 10^{-12}$~cm, $0.363
\cdot 10^{-12}$~cm and $0.5803 \cdot 10^{-12}$~cm for silicon, sodium
and oxygen, respectively. They are taken from Susman
{\it et al.}~(1991) for silicon and oxygen and from 
Bacon (1972) for sodium. Fig.~\ref{fig4} shows 
$S^{{\rm neu}}(q)$ from the simulation and the experiment at 
$T=300$~K. We see that the overall agreement between
simulation and experiment is good. For $q > 2.3$ \AA$^{-1}$
the largest discrepancy is at the peak located at $q=2.8$~\AA$^{-1}$
where the simulation underestimates the experiment by approximately
$15$\% in amplitude. Very well reproduced is the peak at 
$q=1.7$~\AA$^{-1}$. The peak at $q=0.95$ \AA$^{-1}$ is not present 
in the experimental data which might be due to the fact that 
this feature is less pronounced in real systems.

\subsection{Dynamical properties}
\begin{figure}[h]
\psfig{file=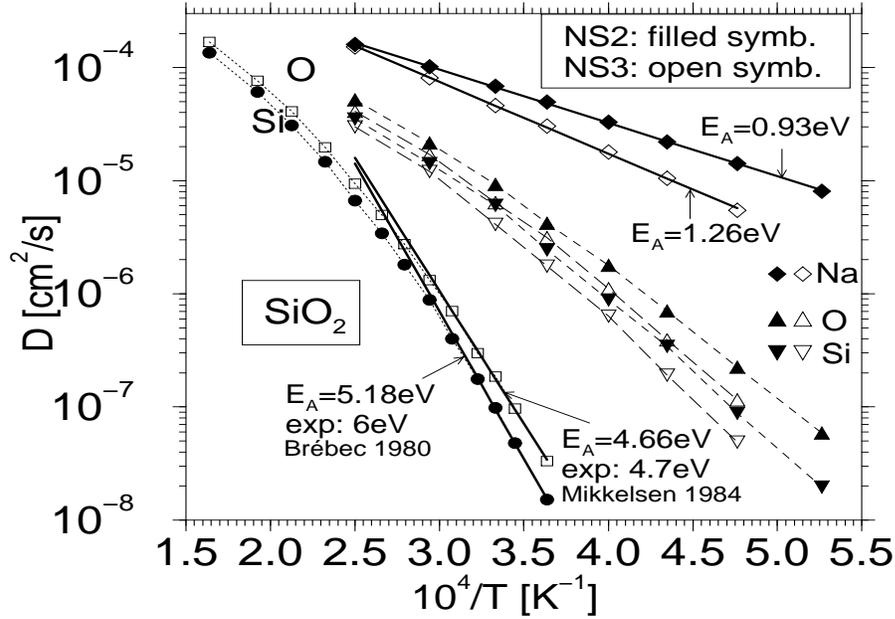,width=13cm,height=9.0cm}
\caption{The diffusion constants for SiO$_2$, NS2, and NS3.
         The bold straight lines are fits with Arrhenius laws.
         The experimental values of the activation energies for
         silica are taken from 
         Br\'ebec {\it et al.}~(1980)
         for silicon and from Mikkelsen (1984) for
         oxygen.}
\label{fig5}
\end{figure}

One of the simplest quantities to study the dynamics of liquids is the
self diffusion constant $D_{\alpha}$ for a tagged particle of type
$\alpha$, which can be calculated from the mean squared displacement
$\left< r_{\alpha}^2 (t) \right>$ via the Einstein relation,
\begin{equation}
  D_{\alpha} = \lim_{t \to \infty} 
	       \frac{\left< r_{\alpha}^2 (t) \right>}{6 t} \ .  
  \label{eq5}
\end{equation}
The different $D_{\alpha}$ for NS2 and NS3 are shown in Fig.~\ref{fig5}
as a function of the inverse temperature. Also included are the
diffusion constants for pure silica from our recent 
simulation (Horbach and Kob, 1999a). For the latter system 
$D_{{\rm Si}}$ and $D_{{\rm O}}$
show at low temperatures the expected
Arrhenius dependence, i.e.~$D_{\alpha}
\propto \exp \left[E_{{\rm A}}/(k_{{\rm B}} T) \right]$. The
corresponding activation energies $E_{{\rm A}}$ are in very good
agreement with the experimental values by Br\'ebec {\it et al.}~(1980)
and Mikkelsen (1984) (given in the figure). We recognize from
Fig.~\ref{fig5} that the dynamics in the sodium silicate melts is much
faster than in SiO$_2$ even at the relatively high temperature
$T=2750$~K ($10^4 / T = 3.64 \; {\rm K}^{-1}$) for which $D_{{\rm Si}}$
and $D_{{\rm O}}$ are about two orders of magnitude larger in NS2 and
NS3 than in SiO$_2$.  Furthermore, we see that the sodium diffusion
constants can be fitted in the whole temperature range very well with
Arrhenius laws.  The corresponding activation energies are $E_{{\rm
A}}=0.93$~eV for NS2 and $E_{{\rm A}}=1.26$~eV for NS3. These
activation energies are about 20 \% to 30 \% higher than those obtained
from electrical conductivity measurements (Greaves and Ngai, 1995, and
references therein). But this discrepancy might at least be partly due
to the fact that our simulations are done at relatively high
pressures.  With decreasing temperature $D_{{\rm Na}}$ decouples more
and more from $D_{{\rm O}}$ and $D_{{\rm Si}}$ in NS2 and NS3.
Therefore, at least at low temperatures, the motion of the oxygen and
silicon atoms is frozen with respect to the timescale of motion
of the sodium atoms.

Experimentally the diffusion constants for oxygen and silicon have been
measured by Poe {\it et al.}~(1997) for the system Na$_2$Si$_4$O$_9$ at
high temperatures and high pressures. Besides a few other combinations
of temperature and pressure they determined $D_{{\rm Si}}$ and $D_{{\rm
O}}$ at $T=2800$~K and $p=10$~GPa where they found the values 
$D_{{\rm Si}} = 1.22 \cdot 10^5$~${\rm cm}^2/{\rm s}$ and 
$D_{{\rm O}} = 1.53 \cdot 10^5$~${\rm cm}^2/{\rm s}$.  
In order to make a comparison with these values we
have done a simulation of Na$_2$Si$_4$O$_9$ for a system of $7680$
particles at the density $2.9$~$g/cm^3$. At this density we have found
at $T=2800$~K and $p=10.13$~GPa the diffusion constants 
$D_{{\rm Si}} = 0.78 \cdot 10^5$~${\rm cm}^2/{\rm s}$, 
$D_{{\rm O}} = 1.17 \cdot 10^5$~${\rm cm}^2/{\rm s}$,
and $D_{{\rm Na}} = 2.35 \cdot 10^5$~${\rm cm}^2/{\rm s}$.  
Thus $D_{{\rm Si}}$ and
$D_{{\rm O}}$ from our simulation underestimate the experimental data
by less than $40$\% which is within the error bars of the experiment.
Therefore it seems that our model gives a quite realistic description
of the diffusion in sodium silicate melts.

\begin{figure}[h]
\psfig{file=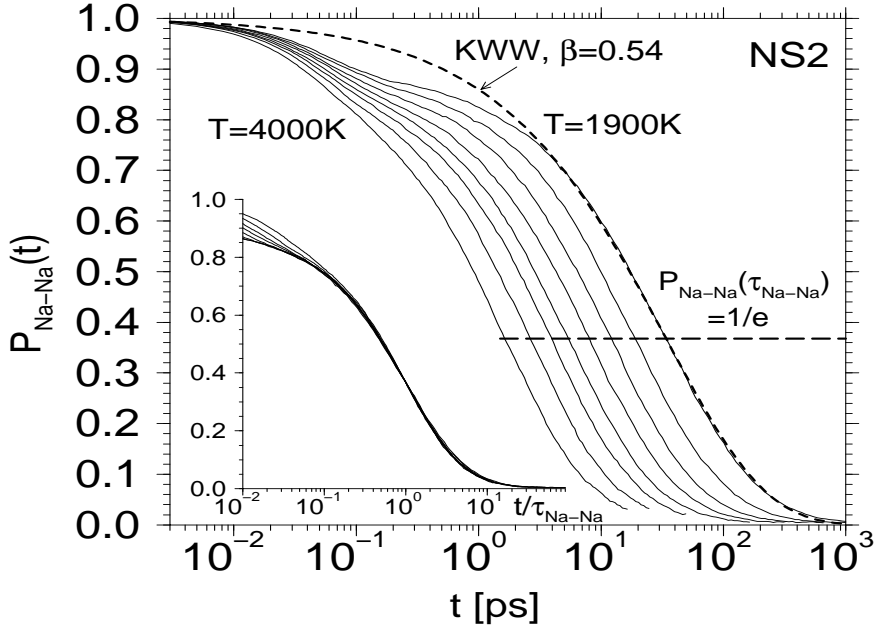,width=13cm,height=9.0cm}
\caption{Time dependence of $P_{{\rm Na-Na}}$, the probability that
         a bond between two sodium atoms which exists at time zero
         is also present at time $t$, for all temperatures 
         investigated. Inset: Plot of the same data versus the
	 scaled time $t/\tau_{{\rm Na-Na}}$ as a function of 
	 temperature.} 
\label{fig6}
\end{figure}

In order to give insight into the microscopic mechanism which is
responsible for the diffusion of the different species in NS2 and NS3
we discuss now the time dependence of the probability $P_{\alpha \beta}
(t)$ that a bond between an atom of type $\alpha$ and an atom of type
$\beta$ is present at time $t$ when it was present at time zero. For
this we define two atoms as bonded if their distance from each other is
less than the location of the first minimum $r_{{\rm min}}$ in the
corresponding partial pair correlation function $g_{\alpha \beta}(r)$.
In the following we restrict our discussion to NS2 because the
conclusions which are drawn below also hold for NS3.
From the functions $g_{\alpha \beta}(r)$ for NS2 we find for
$r_{{\rm min}}$ the values $3.6$~\AA, $5.0$~\AA, $2.35$~\AA, $5.0$~\AA,
$3.1$~\AA, and $3.15$~\AA~for the Si--Si, Si--Na, Si--O, Na--Na, Na--O,
and~O--O correlations. As an example Fig.~\ref{fig6} shows $P_{{\rm
Na-Na}}$ for the different temperatures. First of all we recognize from
this figure that a plateau is formed on an intermediate time scale
which becomes more and more pronounced the lower the temperature is.
This plateau is due to the fact that at $r=r_{{\rm min}}$ the amplitude
of $g_{{\rm Na-Na}}(r)$ is equal $0.56$ and not zero. Thus there are
some sodium atoms which vibrate between the first and the second
neighbor shell leading to the first fast decay of $P_{{\rm Na-Na}}$ to
the plateau value. In the long time regime of $P_{{\rm Na-Na}}$, where
it decays from the plateau to zero, its shape seems to be  independent
of temperature. This is also true for the functions $P_{\alpha
\beta}(t)$ for the other correlations.  For this reason it makes sense
to define the lifetime $\tau_{\alpha \beta}$ of a bond between two
atoms of type $\alpha$ and $\beta$ as the time at which $P_{\alpha
\beta}(t)$ has decayed to $1/{\rm e}$. Indeed, if the function $P_{{\rm
Na-Na}}$ for the different temperatures is plotted versus the scaled
time $t/\tau_{{\rm Na-Na}}$ (inset of Fig.~\ref{fig6}) one obtains one
master curve in the long time regime $t/\tau_{{\rm Na-Na}}>0.1$. This
master curve cannot be described by an exponential function but is well
described by a Kohlrausch--Williams--Watts (KWW) function, $P(t)
\propto \exp(-(t/\tau_{{\rm Na-Na}})^{\beta})$ with an exponent
$\beta=0.54$, which is shown in Fig.~\ref{fig6} in which this function
is fitted to the curve for $T=1900$~K.

\begin{figure}[h]
\psfig{file=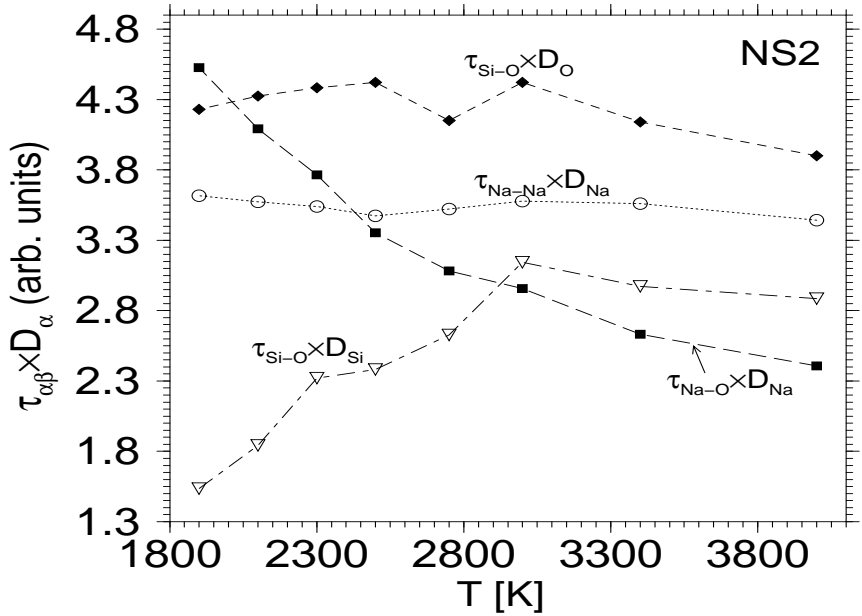,width=13cm,height=9.0cm}
\caption{The products $\tau_{\alpha \beta}\cdot D_{\alpha}$ show
         whether or not the diffusion constant $D_{\alpha}$ is 
         correlated with the lifetime of a bond $\tau_{\alpha\beta}$.}
\label{fig7}
\end{figure}

The lifetimes $\tau_{\alpha \beta}$ can now be correlated with the
diffusion constants by plotting different products $\tau_{\alpha \beta}
\cdot D_{\gamma}$ versus temperature, which is done in Fig.~\ref{fig7}.
The product $\tau_{{\rm Na-Na}} \cdot D_{{\rm Na}}$ is essentially
constant over the whole temperature range whereas $\tau_{{\rm Na-O}}
\cdot D_{{\rm Na}}$ increases with decreasing temperature. This means
that the elementary diffusion step for the sodium diffusion is related
to the breaking of an Na--Na bond and not to that of an Na--O bond,
although the nearest neighbor distance is smaller for Na--O ($r_{{\rm
Na-O}}=2.2$~\AA) than for Na--Na ($r_{{\rm Na-Na}}=3.3$~\AA). We
mention that an Arrhenius law holds also for $\tau_{{\rm Na-O}}$ for
which we have found the activation energy $E_{{\rm A}}=1.14$ eV. The
latter can be interpreted as an effective Na--O binding energy in NS2.
In NS3 $\tau_{{\rm Na-O}}$ can be fitted with an Arrhenius law for
$T<3000$~K, the binding energy in this case is $1.64$ eV.  Also
constant is the product $\tau_{{\rm Si-O}} \cdot D_{{\rm O}}$ which
shows that the oxygen diffusion is related to the breaking of Si--O
bonds. In contrast to that $\tau_{{\rm Si-O}} \cdot D_{{\rm Si}}$ is
only constant at high temperatures. For temperatures $T < 3000$ K this
product decreases with decreasing temperature. Concerning the oxygen
and silicon diffusion we have found recently that the same conclusions
hold also for pure silica (Horbach and Kob, 1999a).

\begin{figure}[h]
\psfig{file=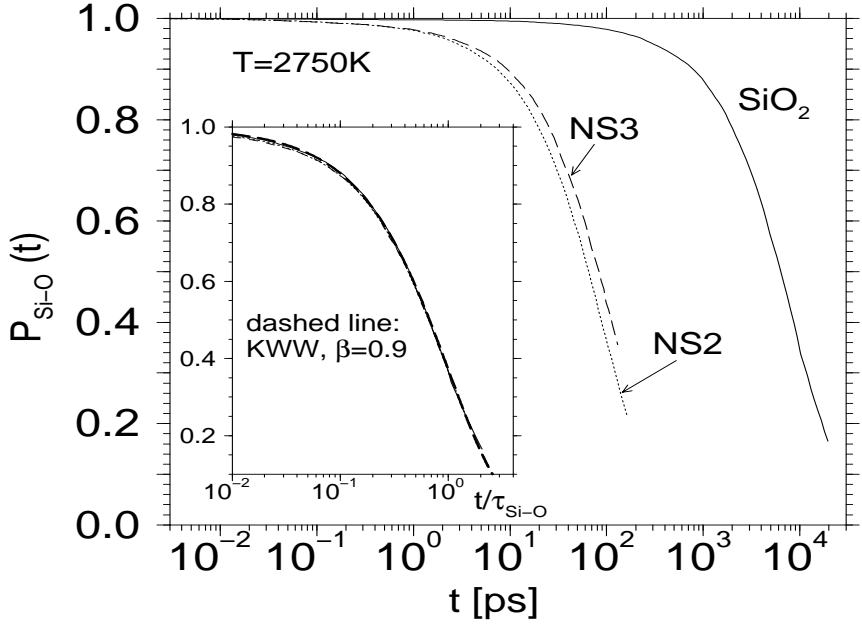,width=13cm,height=9.0cm}
\caption{$P_{{\rm Si-O}}(t)$ for NS2, NS3, and SiO$_2$ at $T=2750$~K.
	 Inset: Plot of the same data versus the scaled time
	 $t/\tau_{{\rm Si-O}}$.}
\label{fig8}
\end{figure}

We have seen that the temperature dependence of $D_{{\rm O}}$ and 
$D_{{\rm Si}}$ for SiO$_2$ changes strongly with the addition
of sodium atoms and, moreover, diffusion becomes much faster.
In Fig.~\ref{fig8} we compare the behavior of the quantity 
$P_{{\rm Si-O}}(t)$ for NS2, NS3 and SiO$_2$ at $T=2750$~K.
We recognize that the shape of the curves seems to be the same for
the three systems. The only difference lies in the time scale which
is, as expected from the behavior of the diffusion constants,
about two orders of magnitudes larger for silica than the sodium
silicate systems. That the shape of the curves is indeed the same
for the three systems is demonstrated in the inset of Fig.~\ref{fig8}
in which we have plotted the same data as before versus the scaled
time $t/\tau_{{\rm Si-O}}$. We see that the three curves fall
nicely onto one master curve which can be fitted by a KWW law with
exponent $\beta=0.9$. This is an astonishing result for $P_{{\rm
Si-O}}(t)$ since the local environment of the oxygen atoms is very
different in the sodium silicate systems from that in silica.

\begin{figure}[h]
\psfig{file=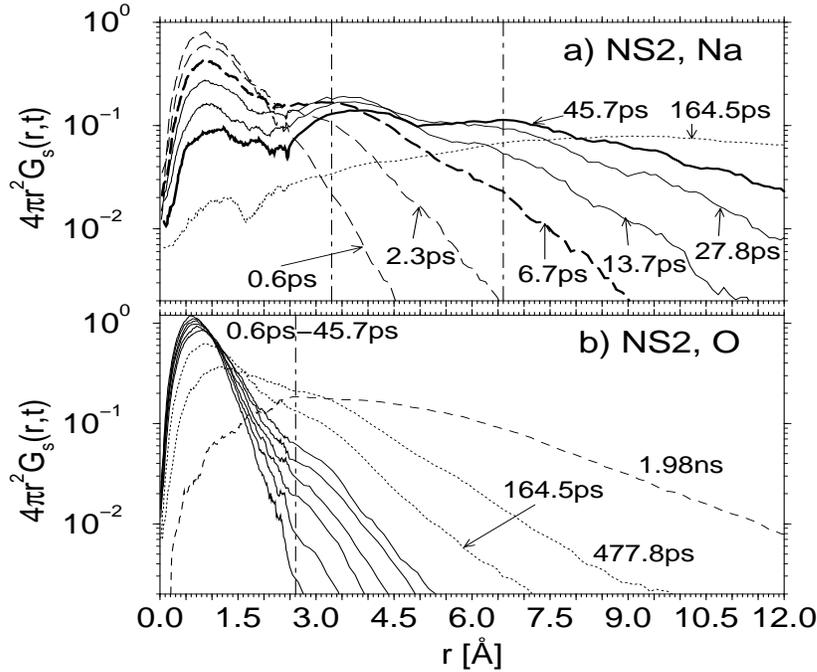,width=13cm,height=9.0cm}
\caption{Space-- and time--dependence of the self part of the van Hove
	 correlation function for NS2 at $T=2100$~K in a
	 linear--logarithmic plot, a) for sodium atoms, b) for
	 oxygen atoms. The vertical lines correspond to
	 $\bar{r}_{{\rm Na-Na}}=3.3$~\AA~and 
	 $2 \bar{r}_{{\rm Na-Na}}$ in a) and 
	 $\bar{r}_{{\rm O-O}}=2.61$ \AA~in b).}
\label{fig9}
\end{figure}

Further insight on how diffusion takes place in sodium silicates can
be gained from the self part of the van Hove correlation function
which is defined by (Balucani and Zoppi, 1994)
\begin{equation}
  G_s^{\alpha} (r,t) = \frac{1}{N_{\alpha}}
	       \sum_{i=1}^{N_{\alpha}} \;
	       \left< \delta(r - | {\bf r}_i (t) - {\bf r}_i (0) | )
	       \right>  \rangle \qquad \alpha \in \{{\rm Si,Na,O}\} 
	       \quad . \label{eq6}
\end{equation}
Thus $4 \pi r^2 G_s^{\alpha}(r,t)$ is the probability to 
find a particle a
distance $r$ away from the place it was at $t=0$. In Figs.~\ref{fig9}a
and \ref{fig9}b we show this probability for different times at
$T=2100$ K for sodium and oxygen, respectively. Note that we have
chosen in both cases a linear--logarithmic plot. At $t=0.6$ ps the
sodium function exhibits a single peak with a shoulder 
around $r>2.8$~\AA. This shoulder becomes more 
pronounced as time goes on and at
$t=6.7$ ps there is a second peak located at a distance $r$ which is
equal to the average distance $\bar{r}_{{\rm Na-Na}}=3.3$ \AA~between
two nearest sodium neighbors (marked with a vertical line in
Fig.~{\ref{fig9}a}). The first peak is still located at the same
position as it was at $t=0.6$ ps while its amplitude has decreased.
Thus we can conclude from this that the sodium atoms do not diffuse in a
continuous way but discontinuously in time by hopping on average over
the distance $\bar{r}_{{\rm Na-Na}}$. This interpretation was first
given for similar features in a soft--sphere system by Roux {\it et
al.}~(1989). At $t=45.7$ ps even a third peak has developed at $2
\bar{r}_{{\rm Na-Na}}$ while the first two peaks remain at the same
position. This means that many sodium atoms have performed now a second
diffusion step. We see from this that two successive elementary
diffusion steps, each corresponding to a breaking of a Na--Na bond, are
spatially highly correlated with each other.  At $t=164.5$ ps the
function has lost its three peak structure but the first peak is still
observable with a significant amplitude.

Whereas many of the sodium atoms have performed two elementary
diffusion steps at $t=45.7$ ps the amplitude of the first peak for
oxygen (Fig.~\ref{fig9}b) decreases only from about 1.5 to 0.8 in the
time interval $0.6 \; {\rm ps} \le t \le 45.7 \; {\rm ps}$. In this
time interval most of the oxygen atoms sit in the cage which is formed
by the neighboring atoms and only rattle around in this cage.
Nevertheless, in this time window a shoulder becomes more and more
pronounced around the mean distance between two nearest oxygen
neighbors $\bar{r}_{{\rm O-O}}=2.61$ \AA. This means that also the
oxygen diffusion takes place by activated hopping events. We have found
the same behavior for oxygen at low temperatures in pure silica
(Horbach and Kob, 1999a).

\section{Summary}
By using a simple pair potential we have performed large scale
molecular dynamics simulations in order to investigate the dynamic
properties of the two sodium silicate melts Na$_2$Si$_2$O$_5$ (NS2) and
Na$_2$Si$_3$O$_7$ (NS3).  The structure of these two systems can be
characterized by a partially destroyed tetrahedral SiO$_4$ network, and
a homogeneous distribution of sodium atoms which have on average about
16 sodium and silicon atoms as nearest neighbors. The regions in
between the network structure introduce a new length scale which is
about two times the distance of nearest Na--Na or Na--Si neighbors.
This leads to a peak in the static structure factor at
$q=0.95$~\AA$^{-1}$. Furthermore, we have found no evidence in the
static structure factor that a microsegregation of Na$_2$O complexes
takes place. A comparison of experimental data to the one of our
computer simulation shows that the latter gives a fair description of
the static properties of NS2 and NS3.  We have explicitly demonstrated
this by showing that the static structure factor $S^{{\rm neu}}(q)$
from a neutron scattering experiment for NS2 (Misawa {\it et al.},
1980) is reproduced quite well by our simulation.  In contrast to our
simulation this experiment exhibits no peak at $q=0.95$~\AA$^{-1}$
which might be due to the fact that this feature is less pronounced in
real systems. Our simulations give not a very good description for the
$Q^{(n)}$ species distribution which is perhaps due to the fact that
our model underestimates the coordination number of nearest Na--O
neighbors.

Nevertheless the oxygen and silicon diffusion constants for
Na$_2$Si$_4$O$_9$ are in very good agreement with experimental data by
Poe {\it et al.}~(1997). In comparing the diffusion constants in NS2
and NS3 with those in silica we have recognized that the dynamics
becomes much faster with the addition of the sodium atoms. Moreover, in
the sodium silicates the diffusion constant for sodium decouples for
decreasing temperature more and more from those of oxygen and silicon,
such that at low temperatures the dynamics of the silicon and oxygen
atoms is frozen in with respect to the movement of the sodium atoms.
The sodium diffusion constants for NS2 and NS3 exhibit over the whole
temperature range we investigate an Arrhenius behavior with activation
energies around $1$~eV.

We have shown that our simulation is able to give insight into the
microscopic mechanism of diffusion in NS2 and NS3. From the time
dependent bond probability $P_{\alpha \beta}(t)$ we determined an
average lifetime $\tau_{\alpha \beta}$ of bonds between an atom of type
$\alpha$ and an atom of type $\beta$.  By correlating the lifetimes
$\tau_{\alpha \beta}$ with the diffusion constants, we show that the
elementary diffusion step in the case of sodium is the breaking of a
Na--Na bond and in the case of oxygen that of a Si--O bond. By studying
the self part of the van Hove correlation function we have demonstrated
that at low temperatures the diffusion for oxygen and sodium takes
place by activated hopping events over the mean distance $\bar{r}_{{\rm
O-O}}=2.6$~\AA~and $\bar{r}_{{\rm Na-Na}}=3.3$~\AA, respectively.
Moreover, we have found for the case of sodium that at least two
successive diffusion steps are spatially highly correlated with each
other.

Acknowledgments:
We thank K.--U. Hess, D. Massiot, B. Poe, and J. Stebbins for useful
discussions on this work. This work was supported by SFB 262/D1 and by
the Deutsche Forschungsgemeinschaft, Schwerpunktsprogramm 1055. We
thank the HLRZ Stuttgart for a generous grant of computer time on the
CRAY T3E.

\section{References}

\begin{trivlist}
\item[]
Angell, C. A., Clarke, J. H. R., and Woodcock, L. V., 1981. 
Interaction potentials and glass formation, a survey of computer
experiments. Adv. Chem. Phys., 48: 397--453.
\item[]
Bacon, G. E., 1972. Acta Cryst. A, 28: 357.
\item[]
Balucani, U., and Zoppi, M., 1994. Dynamics of the Liquid State. 
Clarendon Press, Oxford.
\item[]
Br\'ebec, G., Seguin, R., Sella, C., Bevenot, J., and Martin, J. C.,
1980. Diffusion du silicium dans la silice amorphe.
Acta Metall., 28, 327--333.
\item[]
Brown, G. E., Farges, F., Calas, G., 1995. X--Ray Scattering and
X--Ray Spectroscopy Studies of Silicate Melts. 
Rev. Mineral., 32: 317--410.
\item[]
Cormack, A. N., and Cao, Y., 1997. Molecular Dynamics Simulation of
Silicate Glasses. In: B. Silvi and P. Arco (Eds.), Modelling of 
Minerals and Silicated Materials, Kluwer, Dordrecht: 227--271.
\item[]
Greaves, G. N., and Ngai, K. L., 1995. Reconciling ionic--transport
properties with atomic structure in oxide glasses. Phys. Rev. B, 52:
6358--6380.
\item[]
Haller, W., Blackburn, D. H., and Simmons, J. H., 1974. Miscibility
gaps in alkali--silicate binaries --- data and thermodynamic
interpretation. J. Am. Ceram. Soc., 57: 120--126.
\item[]
Horbach, J., Kob, W., and Binder, K., 1996. Finite size effects in
simulations of glass dynamics. Phys. Rev. E, 54: R5897--R5900.
\item[]
Horbach, J., and Kob, W., 1999a. Static and Dynamic Properties of a
Viscous Silica Melt. Phys. Rev. B 60.
\item[]
Horbach, J., Kob, W., and Binder, K., 1999b. High frequency dynamics
of amorphous silica. Submitted to Phys. Rev. B.
\item[]
Horbach, J., and Kob, W., 1999c. The Structure and Dynamics of 
Sodium Disilicate. Submitted to Phil. Mag. B.
\item[]
Knoche, R., 1993. Temperaturabh\"angige Eigenschaften silikatischer 
Schmelzen. Ph. D. thesis, Bayreuth: 100--109.
\item[]
Knoche, R., Dingwell, D. B., Seifert, F. A., and Webb, S. L., 1994.
Non--linear properties of supercooled liquids in the system
Na$_2$O--SiO$_2$. Chem. Geol., 116: 1--16.
\item[]
Kob, W., 1999. Computer simulations of supercooled liquids and glasses.
J. Phys.: Condens. Matter, 11: R85--R115.
\item[]
Kramer, G. J., de Man, A. J. M., and van Santen, R. A., 1991. 
Zeolites versus Aluminosilicate Clusters: The Validity of a Local
Description. J. Am. Chem. Soc., 64: 6435--6441.
\item[]
Mazurin, O. V., Kluyev, V. P., and Roskova, G. P., 1970. The influence
of heat treatment on the viscosity of some phase separated glasses.
Phys. Chem. Glass., 11: 192--195.
\item[]
McMillan, P. F., and Wolf, G. H., 1995. Vibrational spectroscopy
of silicate liquids. Rev. Mineral., 32: 247--315.
\item[]
Mikkelsen, J. C., 1984. Self--diffusivity of network oxygen in
vitreous SiO$_2$. Appl. Phys. Lett., 45: 1187--1189.
\item[]
Misawa, M., Price, D. L., and Suzuki, K., 1980. 
The short--range structure of alkali disilicate glasses by 
pulsed neutron total scattering.
J. Non--Cryst. Solids, 37: 85--97.
\item[]
Mysen, B. O., and Frantz, J. D., 1992. Raman spectroscopy of 
silicate melts at magmatic temperatures: Na$_2$O--SiO$_2$,
K$_2$O--SiO$_2$ and LiO$_2$--SiO$_2$ binary compositions in the
temperature range 25--1475 $^{\circ}{\rm C}$. Chem. Geol., 96:
321--332.
\item[]
Poe, B. T., McMillan, P. F., Rubie, D. C., Chakraborty, S., Yarger, J.,
and Diefenbacher, J., 1997. Silicon and Oxygen Self--Diffusivities
in Silicate Liquids Measured to 15 Gigapascals and 2800 Kelvin.
Science, 276: 1245--1248.
\item[]
Roux, J. N., Barrat, J. L., and Hansen, J.--P., 1989. Dynamical
diagnostics for the glass transition in soft--sphere alloys.
J. Phys.: Condens. Matter, 1: 7171--7186.
\item[]
Smith, W., Greaves, G. N., and Gillan, M. J., 1995. Computer simulation
of sodium disilicate glass. J. Chem. Phys., 103: 3091--3097.
\item[]
Sprenger, D., Bach, H., Meisel, W., and G\"utlich, P., 1992.
Discrete bond model (DBM) of binary silicate glasses derived from
${}^{29}$Si--NMR, Raman, and XPS measurements. In: The Physics of
Non--Crystalline Solids (Eds.: D. L. Pye, W. C. La Course, and 
H. J. Stevens), 42--47.
\item[]
Stebbins, J. F., 1988. Effects of temperature and composition on 
silicate glass structure and dynamics: Si--29 NMR results.
J. Non--Cryst. Solids, 106: 359--369.
\item[]
Stebbins, J. F., 1995. Dynamics and Structure of Silicate and
Oxide Melts: Nuclear Magnetic Resonance Studies. 
Rev. Mineral., 32: 191--247.
\item[]
Susman, S., Volin, K. J., Montague, D. G., and Price, D. L., 1991.
Temperature dependence of the first sharp diffraction peak in 
vitreous silica. Phys. Rev. B, 43: 11076--11081.
\item[]
van Beest, B. W. H., Kramer, G. J., and van Santen, R. A., 1990.
Force Fields for Silicas and Aluminophosphates Based on {\it Ab Initio}
Calculations. Phys. Rev. Lett., 64: 1955--1958.
\item[]
Vessal, B., Amini, M., Fincham, D., and Catlow, C. R. A., 1989.
Water--like Melting Behavior of SiO$_2$ Investigated by the Molecular
Dynamics Simulation Techniques. Philos. Mag. B, 60: 753--775.
\item[]
Vessal, B., Greaves, G. N., Marten, P. T., Chadwick, A. V., Mole, R.,
and Houde--Walter, S., 1992. Cation microsegregation and ionic
mobility in mixed alkali glasses. Nature, 356: 504--506.
\end{trivlist}
%
%\begin{figure}[h]
%\caption{Life time $\tau_{{\rm Si-O}}$ as a function of the Na$_2$O
%         concentration at different temperatures. The lines are just
%         guides for the eye.}
%\label{fig9}
%\end{figure}
%
\end{document}